\documentclass[journal]{IEEEtran}
\usepackage{xcolor}
\usepackage[colorlinks=true, linkbordercolor=white]{hyperref}

\ifCLASSINFOpdf
  \usepackage[pdftex]{graphicx}
\else
\fi
\usepackage{amsmath}
\usepackage{caption}
\usepackage{subcaption}
\usepackage{algorithm}
\usepackage{algpseudocode}
\usepackage{svg}

\usepackage{amssymb}
\usepackage{mathtools}
\usepackage{amsthm}
\usepackage{multicol}
\usepackage{adjustbox}

\usepackage{calc}
\usepackage{lscape}
\usepackage{afterpage}

\usepackage{array}
\begin{document}
%
\title{Explainable AI in Speaker Recognition - Making Latent Representations Understandable} 
%
%
%

\author{Yanze Xu, 
        Wenwu Wang,~\IEEEmembership{Fellow,~IEEE,}
        and~Mark D. Plumbley,~\IEEEmembership{Fellow,~IEEE}
\thanks{Yanze Xu, Wenwu Wang are in the Centre for Vision, Speech and Signal Processing, University of Surrey, Guildford, GU2 7XH, UK. Mark D. Plumbley is in Department of Informatics, King's College London, London, WC2R 2LS, UK. Corresponding author: Yanze Xu [yanze.xu@outlook.com]}}

\maketitle

\begin{abstract}
Neural networks can be trained to learn task-relevant representations from data. Understanding how these networks make decisions falls within the Explainable AI (XAI) domain. This paper proposes to study an XAI topic: analysing, visualising and understanding the unknown organisation of network representations, particularly those a speaker recognition network learns from utterances, for recognising speaker identity.

Past studies have employed algorithms (e.g. K-means) to analyse the different ways in which network representations can be naturally grouped into clusters, i.e. to analyse different flat clustering phenomena within the space defined by those representations. In contrast, this work applies two algorithms---Single-Linkage Clustering (SLINK) and Hierarchical Density-Based Spatial Clustering of Applications with Noise (HDBSCAN)---to analyse the different ways in which representations from the speaker recognition network can form clusters with hierarchical relationships, i.e., to analyse different hierarchical clustering phenomena within the representation space of the speaker recognition network.

Furthermore, an algorithm called Hierarchical Cluster-Class Matching (HCCM) is designed to semantically interpret one of the above hierarchical clustering phenomena analysed using SLINK. Given the clusters representing this phenomenon, HCCM identifies which ones best match individual semantic classes related to gender and nationality (e.g.\ male, female, Ireland, UK) and and-logic conjunctions of these classes (e.g.\ female and Ireland). The Liebig score metric is also proposed within HCCM to quantify the matching quality of each cluster-class pair and diagnose the factor that limits each match.

\end{abstract}


%
\IEEEpeerreviewmaketitle

\section{Introduction} \label{sec:intro}

\IEEEPARstart{E}xplainable AI (XAI) aims to make the decision-making processes of AI systems, particularly those implemented using neural networks~\cite{he2016deep, lecun2015deep}, explainable and understandable to humans~\cite{gunning2019darpa, xu2019explainable, linardatos2020explainable}. In the deep learning field~\cite{he2016deep, lecun2015deep}, neural networks are typically trained in a supervised manner to learn a nonlinear mapping from inputs (e.g. images or signals) to pre-labelled outputs (e.g. classes or values), with the intermediate outputs of this mapping process serving as task-relevant representations of the inputs~\cite{nasteski2017overview}.

We are interested in studying two high-level XAI questions: \textbf{i)} Inspired by the fact that humans can organise knowledge or information~\cite{lynn2020humans}, we assume that the neural network has a similar ability to organise representations of different inputs, and we question what the network’s organisation of representations is. \textbf{ii)} Inspired by the fact that human attention mechanism is the cognitive ability to selectively process relevant stimuli~\cite{johnston1986selective}, we assume that the network has the computational ability to selectively process task-relevant information during decision-making, and we question what information is selectively processed. The second question is studied in our other work~\cite{xu2026explainableaispeakerrecognition}, and this paper focuses exclusively on the first question. To address this, we formally propose to analyse, visualise, and understand the unknown organisation within the multi-dimensional space defined by representations extracted from a well-trained neural network, the \textit{network representation space}. Our experiments are conducted within the speaker recognition task, where most speaker recognition neural networks are trained to map an utterance or its spectrogram to a pre-labelled speaker identity~\cite{cai2018exploring, nagrani2017voxceleb, chung2020defence}.


\begin{figure}
    \centering
    \includegraphics[width=0.99\linewidth]{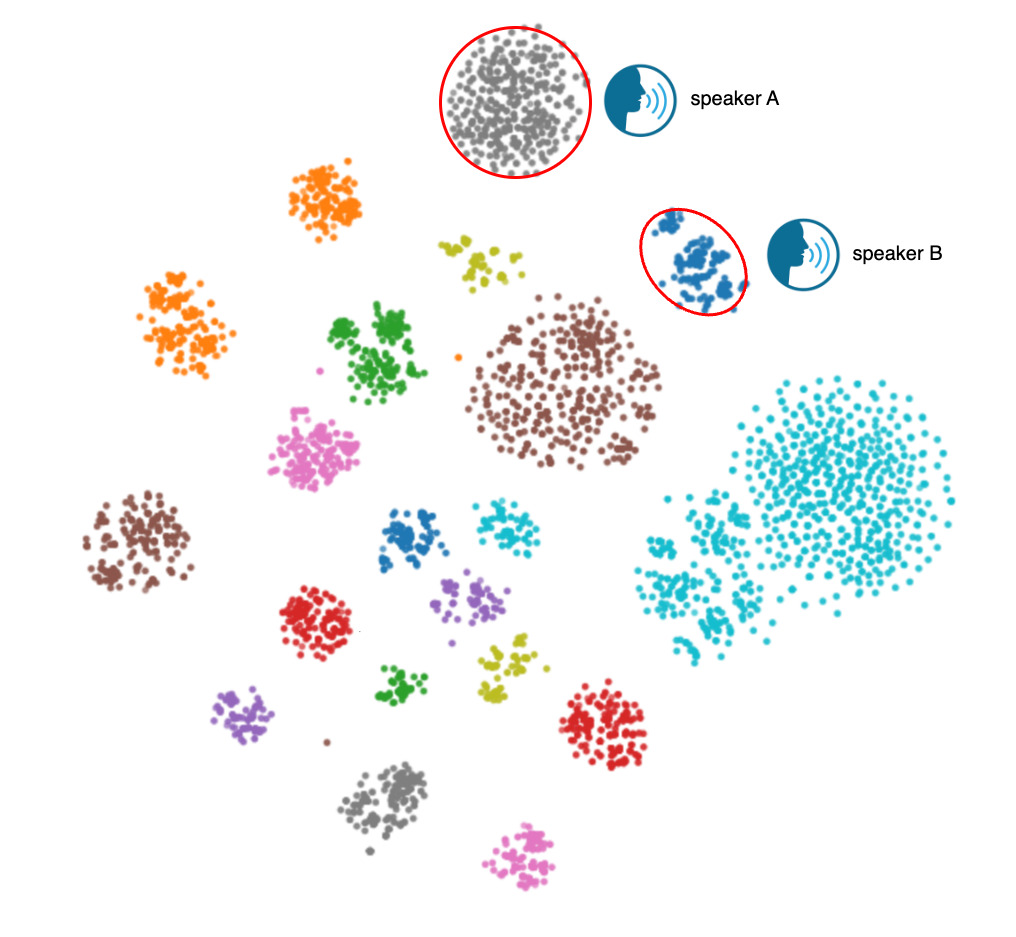}
    \caption{A 2-dimensional visualisation approximating the representation space of a well-trained speaker recognition network, originated from Li et al.'s paper~\cite{li2024efficient}.}
    \label{fig:tSNE_embed_space}
\end{figure}


Some works employ flat clustering algorithms~\cite{manning2008flat}, such as K-means~\cite{macqueen1967some}, to analyse how network representations are naturally organised into independent, flattened clusters, where different algorithms reveal different natural organisations of representations~\cite{Peiffer2021, carbonnelle2020intraclass, caron2018deep}. We define each natural organisation of representations into flattened clusters as a certain flat clustering phenomenon happening inside the space formed by these representations, which Peiffer~\cite{Peiffer2021} referred to as \textit{inner} or \textit{internal clustering} phenomenon for short. Moreover, other studies~\cite{prasad2024tree, tawara2020frame, li2024efficient} use dimensionality reduction algorithms~\cite{van2008visualizing, mcinnes2018umap, prasad2024tree} to visualise network representations within an approximated low-dimensional space, where inner clustering phenomena can be intuitively observed. As shown in Fig.~\ref{fig:tSNE_embed_space}, Li et al.~\cite{li2024efficient} present a 2-dimensional space that approximates the representation space of their well-trained speaker recognition network, in which representations are indeed organised into flat clusters. To understand the inner clustering phenomenon in Fig.~\ref{fig:tSNE_embed_space}, Li et al.\ interpret that each flat cluster corresponds to a particular speaker identity.

However, to our limited knowledge, the above studies on organisations of network representations have been largely discussed outside the XAI domain~\cite{xu2019explainable, 9369420, linardatos2020explainable}, motivating us to study this topic from an XAI perspective. Moreover, the XAI domain still lacks studies systematically explaining how network representations can be naturally and unbiasedly organised into clusters with deeper relationships, rather than treating them as independent. In this work, we aim to employ hierarchical clustering algorithms~\cite{ran2023comprehensive} to analyse whether hierarchical relationships remain in representation clusters in the speaker recognition task. We define each way (e.g. each algorithm) of organising network representations into hierarchical clusters as a certain hierarchical clustering phenomenon within the corresponding network representation space, referred to as an \textit{inner hierarchical clustering} phenomenon for short.

To achieve our purpose, two well-established hierarchical clustering algorithms, Single-Linkage Clustering (SLINK)~\cite{gower1969minimum, sibson1973slink} and Hierarchical Density-Based Spatial Clustering of Applications with Noise (HDBSCAN)~\cite{campello2013density}, are applied separately to representations extracted from a speaker recognition network trained by Chung et al.~\cite{chung2020defence}. The experiments reveal different inner hierarchical clustering phenomena for the examined speaker recognition network, with SLINK and HDBSCAN each producing hierarchical representation clusters. The quality of each inner hierarchical clustering phenomenon is evaluated using the method proposed by Rosenberg et al.~\cite{rosenberg2007v}.

Afterwards, our research purpose shifts to visualising and understanding the hierarchical clustering phenomena within the representation space of the speaker recognition network. Following the previous experiments, we further visualise the inner hierarchical clustering phenomenon analysed by SLINK, whose hierarchical representation clusters achieve the best evaluation quality under Rosenberg et al.'s method~\cite{rosenberg2007v}, in the form of a tree-like structure, namely a dendrogram~\cite{sokal1958statistical}. Moreover, our additional experiments provide a semantic interpretation for the clusters in the visualised dendrogram using a new method we proposed, termed Hierarchical Cluster-Class Matching (HCCM). In the remaining paragraphs of this section, we give an introduction to HCCM.

\begin{figure}
    \centering
    \includegraphics[width=1\linewidth, height=135pt]{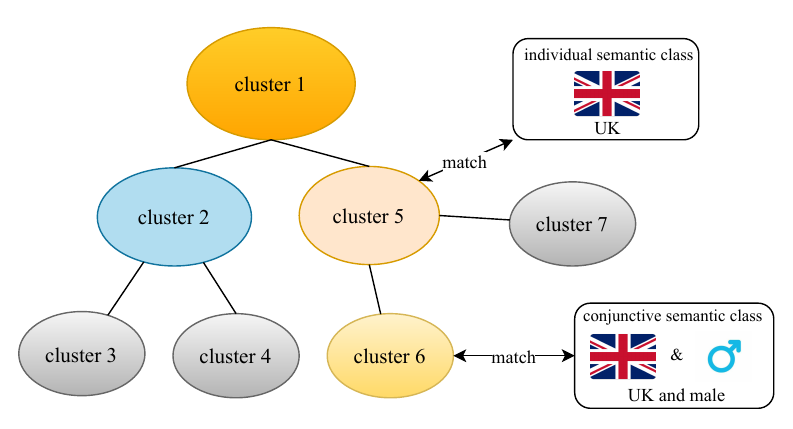}
    \caption{An illustration of the proposed Hierarchical Cluster-Class Matching (HCCM) method for interpreting hierarchical clusters.}
    \label{fig:HCCM_intro}
\end{figure}

HCCM requires network representations to be pre-labelled into classes, where representations of the same class form a \textit{predefined representation division}. On this bais, HCCM performs pairwise matching between hierarchical representation clusters and predefined representation divisions, identifying the best-matched representation cluster for each predefined representation division. As shown in Fig.~\ref{fig:HCCM_intro}, HCCM interprets `cluster 5' as the UK class, considering representations inside `cluster 5' best match with the predefined representation division of the UK class. `cluster 6' is interpreted as the UK\&male semantic class, since representations inside `cluster 6' best match the predefined division whose representations belong to both UK and male classes simultaneously. HCCM refers to the semantic class serving as an individual modifier (e.g. UK) as the \textit{individual semantic class}, and those forming a conjunction~\cite{morzycki2016modification, boole1854investigation} of different individual semantic classes based on and-logic (e.g. UK\&male) as the \textit{conjunctive semantic class}.



\begin{figure}
    \centering
    \includegraphics[width=1\linewidth]{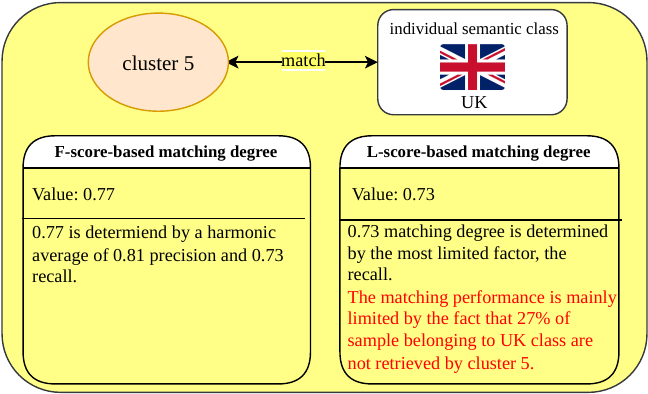}
    \caption{A comparison of matching scores quantified by F-score and L-score, respectively, in the HCCM method.}
    \label{fig:FvsLscore}
\end{figure}

Notably, HCCM provides a new metric, termed the \textit{Liebig score} (L-score), to quantify the matching degree, instead of using the conventional F-score, which is determined by the harmonic mean of recall and precision. The design of L-score is inspired by Liebig's law of the minimum~\cite{de1994liebig}, which states that the overall performance of a system is limited by its most restrictive factor. As illustrated in the right panel of Fig.~\ref{fig:FvsLscore}, a 0.73 L-score-based matching degree is directly determined by the limiting factor, i.e., the 0.73 recall, interpreted as the 27\% of samples belonging to the UK class that are not retrieved by `cluster 5'. In contrast, the left panel of Fig.~\ref{fig:FvsLscore} shows that a 0.77 F-score-based matching degree is determined by the harmonic mean of the 0.73 recall and 0.81 precision, but what this harmonic mean represents is hard to interpret~\cite{christen2023review}.

\section{Background and Related works} \label{sec:related}

\subsection{XAI in Recognition and Generation}
XAI research spans not only neural networks trained for recognition tasks, but also those trained for generative tasks, albeit with different emphases. XAI research for recognition tasks tends to be more theoretical in nature, focusing on explaining attention mechanisms~\cite{lundberg2017unified, ribeiro2016should, selvaraju2016grad, zhou2016learning, petsiuk2018rise}, forgetting mechanisms~\cite{ede2022explain}, and representation organisation. However, these theoretical findings have not translated into widespread practical applications. In contrast, XAI research for generative tasks addresses not only theoretical issues, such as hallucinations~\cite{cambria2024xaimeetsllmssurvey} and emergent abilities~\cite{wei2022emergent}, but has also yielded work of practical value. For instance, some studies identify vectors that capture the principal variance across representations extracted from generative networks~\cite{shen2021closed, 9578611, goetschalckx2019ganalyze}. By modifying representations along such vectors, the generated outputs can be progressively controlled in specific semantic directions, enabling direct manipulation of real-world generative systems.

\subsection{Flat clustering and dimensionality reduction algorithms for network representations} Flat clustering methods~\cite{manning2008flat} and dimensionality reduction methods~\cite{espadoto2019toward} can separately analyse or visualise inner clustering phenomena of neural networks~\cite{Peiffer2021, carbonnelle2020intraclass, caron2018deep, prasad2024tree, tawara2020frame, li2024efficient}. Specifically, a flat clustering algorithm analyses how original high-dimensional representations form flat clusters in a certain way, whereas a dimensionality reduction method transforms these high-dimensional representations into lower-dimensional ones that can be visualised. Although a flat clustering phenomenon may appear within the visualised low-dimensional representation space, it remains uncertain whether the same flat clustering phenomenon preserve in the original representation space.

\subsection{Hierarchical clustering algorithms for network representations} Hierarchical clustering algorithms~\cite{murtagh2017algorithms} can analyse the inner hierarchical clustering phenomena of neural networks. Specifically, using different hierarchical clustering algorithms, Naumov et al.~\cite{Naumov_Yaroslavtsev_Avdiukhin_2021} analyse multiple hierarchical clustering phenomena within the representation space of networks trained for image classification and Natural Language Processing (NLP) tasks, where each phenomenon reveals a group of representation clusters with hierarchical relationships. Similar analyses have also been conducted on speaker recognition networks, where the hierarchical representation clusters each algorithm produced can be used to build speaker diarisation systems addressing the `who speaks when' problem~\cite{garcia2017speaker, singh2023supervised, lukic2016speaker}. Unlike these works, we not only identify multiple inner hierarchical clustering phenomena in our speaker recognition network, but also visualise one of these phenomena as a dendrogram and interpret its corresponding hierarchical representation clusters using the proposed Hierarchical Cluster-Class Matching (HCCM) method.



\subsection{Cluster-class matching (CCM) and HCCM}~\label{sec:CCM_quick_look}

The HCCM method we designed is inspired by a similar approach proposed by Rosenberg et al.~\cite{rosenberg2007v}. Specifically, Rosenberg et al.'s method, also termed the Cluster-Class Matching (CCM) method, is originally designed to measure a global matching score, representing how well independent data clusters produced by a flat clustering algorithm match the predefined data groups (i.e. divisions) of different classes. A better flat clustering algorithm is expected to analyse a group of flat data clusters that align more closely with predefined data divisions. In contrast, HCCM extends this idea in two key ways. First, instead of flat data clusters, HCCM operates on hierarchical representation clusters. Second, instead of computing an overall matching score, HCCM aims to find pairwise matches between predefined representation divisions and hierarchical representation clusters, identifying the best-matched cluster for each predefined division.

\section{Preliminary knowledge}
This work applies two hierarchical clustering algorithms~\cite{ran2023comprehensive} to analyse representations from the speaker recognition neural network: Single-Linkage Clustering (SLINK)~\cite{gower1969minimum, sibson1973slink} and Hierarchical Density-Based Spatial Clustering of Applications with Noise (HDBSCAN)~\cite{campello2013density}. Section~\ref{sec:DBSCAN} introduces the Density-Based Spatial Clustering of Applications with Noise (DBSCAN)~\cite{ester1996density} algorithm, which provides the preliminary knowledge for understanding HDBSCAN and shares a relationship with SLINK. In addition, Section~\ref{sec:old_ccm_method} introduces the Cluster-Class Matching (CCM) method proposed by Rosenberg et al.~\cite{rosenberg2007v}, which serves as the foundation for our proposed Hierarchical Cluster-Class Matching (HCCM) that is designed to semantically interpret hierarchical representation clusters.

\subsection{DBSCAN} \label{sec:DBSCAN}

\begin{figure} 
   \centering
   \includegraphics[width=0.5\textwidth]{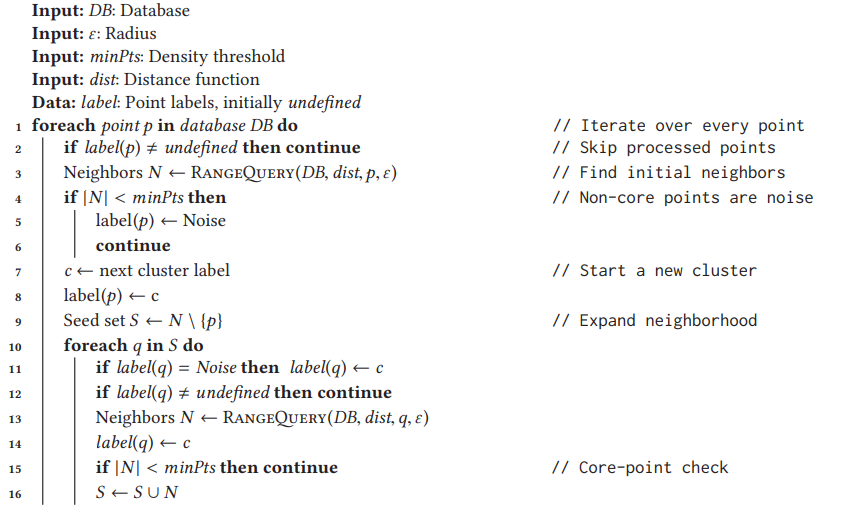}
  \caption{Pesudocode of the DBSCAN from RJGB Campello's work~\protect\cite{campello2013density}.} \label{fig:pesudo}
\end{figure}

DBSCAN~\cite{ester1996density} is one of the most well-known flat clustering algorithms~\cite{xu2015comprehensive} that can analyse flat clustering phenomena present in the data, where each phenomenon represents a natural organisation of data as flat clusters. It is commonly referred to as a density-based flat clustering method, since it captures such natural organisations through a density constraint, requiring each data point assigned to a cluster to have a sufficient number of neighbouring points within a specified search radius.


We now briefly discuss DBSCAN with reference to the pseudocode presented in the original paper~\cite{campello2013density}, as shown in Fig.~\ref{fig:pesudo}. This pseudocode can be divided into two stages. In the first stage (i.e. lines $2\sim9$), the algorithm selects an unvisited data point $p$ as the query point and calls \textsc{RangeQuery()} function in line 3, which takes four parameters: $DB$, the set of all data points; $dist$, the distance metric (typically Euclidean distance); $p$, the query point itself; and $\varepsilon$, the search radius. \textsc{RangeQuery()} returns all data points within the search radius $\varepsilon$ for the query point. If the number of returned points is at least $minPts$, then $p$ is referred to as a \textit{core point}~\cite{ester1996density}, its neighbours within $\varepsilon$ are stored in the set $S$, and $p$ is identified as a valid start point for a new cluster.

In the second stage (i.e. lines $10\sim16$), the algorithm iterates through each point in $S$. For each unvisited point, \textsc{RangeQuery()} is called again to check whether it qualifies as a core point. If so, its neighbours are added to $S$, further expanding the cluster. All core points discovered in this iterative process are labelled as members of the current cluster using $label()$. When all data points in $S$ are visited and no new members can be added, all memberships of the current cluster are labelled. The algorithm then selects the next unvisited and unlabelled point and repeats both stages to form a new cluster, until all data points in $DB$ have been visited, finally yielding a group of independent flat clusters.

\subsection{Cluster-Class Matching for evaluation}~\label{sec:old_ccm_method}

Diverse flat clustering algorithms have been developed to analyse flat clustering phenomena in data, necessitating the evaluation of which algorithm analyses the best flat clustering phenomenon, or in other words, which algorithm produces flat clusters of better quality. To do so, Rosenberg et al.~\cite{rosenberg2007v} proposed a method to evaluate how well flat data clusters analysed by a certain algorithm overall align with the predefined grouping information of the data, referring to this evaluation as cluster-class matching (CCM). In more detail, the predefined grouping information of the data comes from pre-labelling data with different semantic classes, such that all data belonging to the same class form a group, referring to as a \textit{predefined data division}. On this basis, CCM computes a matching score representing the overall alignment between data clusters and predefined data divisions. A better flat clustering algorithm is expected to achieve a higher overall matching score, meaning that the flat clustering phenomenon analysed by this algorithm aligns more closely with the predefined grouping information. Subsequent studies~\cite{johnson2013comprehensive, zhao2005hierarchical} generalised the CCM method to evaluate hierarchical clustering algorithms instead of flat clustering algorithms, using the F-score~\cite{christen2023review} or V-score~\cite{rosenberg2007v} to quantify the overall matching score.

We now present the mathematical formulation of the CCM method in detail. Specifically, let $C = \{c_1, c_2, \dots, c_{n_\mathrm{class}}\}$ denote the collection of predefined data divisions of $n_\mathrm{class}$ semantic classes, where each $c_i$ is a set that gathers the indices of all data points whose groundtruth labels belong to the $i$-th semantic class. In addition, let $K = \{k_1, k_2, \dots, k_{n_\mathrm{cluster}}\}$ denote the collection of $n_\mathrm{{{cluster}}}$ data clusters analysed by a certain flat clustering algorithm, where each $k_j$ set contains the indices of all data points assigned to the $j$-th cluster. CCM calculates the overall matching degree between data clusters $K$ and predefined data divisions $C$ as follows:

\begin{equation}
F_{\mathrm{score}}(C, K) = \sum_{c_i \in C} \frac{|c_i|}{\mathcal{N}} \max_{k_j \in K} f_{\mathrm{score}}(c_i, k_j) \label{eq:Fscore}
\end{equation}

\noindent Here, $|c_i|$ represents the number of data points pre-labelled as the $i$-th semantic class, and $\mathcal{N}$ is the total number of data points in the dataset; thus, $|c_i| / \mathcal{N}$ specifies the proportion of data points belonging to the $i$-th semantic class. Given the predefined data division of the $i$-th class (i.e. $c_i$), the expression $\max_{k_j \in K} f_{\mathrm{score}}(c_i, k_j)$ iterates over all clusters in $K$ to find the highest matching score between $c_i$ and any cluster in $K$, as quantified by the F-score metric $f_{\mathrm{score}}$. The highest matching scores of $n_\mathrm{class}$ predefined data divisions are then combined as a weighted average to produce the overall matching score. A higher overall matching score indicates that the data clusters $K$ exhibit stronger alignment with the predefined grouping information (i.e. $C$), reflecting a superior clustering phenomenon analysed by the algorithm. Moreover, The F-score metric \( f_\mathrm{score} \) in Equation~\eqref{eq:Fscore}, is defined as follows:

\begin{equation*}
\begin{aligned}
f_{\mathrm{score}}(c, k) 
    = \frac{2}{\left(\dfrac{|c \cap k|}{|c|}\right)^{-1} 
        + \left(\dfrac{|c \cap k|}{|k|}\right)^{-1}}
    = \frac{2|c \cap k|}{|c| + |k|}
    ~\label{eq:fscore1}
\end{aligned}
\end{equation*}

\noindent Here, $\frac{|c \cap k|}{|c|}$ calculates the recall value for matching the predefined data division of a certain class with a certain data cluster. The recall in this context is defined as the proportion of all data points belonging to the semantic class that are successfully retrieved by the matched cluster~\cite{rosenberg2007v}. In addition, $\frac{|c \cap k|}{|k|}$ is the precision, which is defined as the proportion of data points in the cluster that truly belong to, or are predicted as, the matched semantic class~\cite{rosenberg2007v}. On this basis, the F-score value is a harmonic mean of the values of precision and recall.

However, Christen et al.~\cite{christen2023review} criticised the fact that, although both precision and recall are proportions with clear representational meaning, their harmonic mean (i.e. the F-score) lacks such interpretability. As an example, in the context of a document retrieval task, a recall of 0.30 can be directly interpreted as `30\% of the relevant documents are retrieved by the system', while a precision of 0.30 indicates that `30\% of the retrieved documents are relevant'~\cite {manning2008introduction}. By contrast, an F-score value of 0.30 cannot be interpreted directly: it indicates a balance of precision and recall, but cannot be interpreted as any kind of proportion of documents.

\begin{figure*}
\begin{minipage}[t]{0.5\textwidth}
\begin{algorithm}[H]
    \centering
    \caption{Find a flat cluster by DBSCAN \textcolor{white}{(asdfasdfasdfasdfasdfasdf)}}\label{algorithm1}
    \begin{algorithmic}[1]
    \State $N \gets \textsc{RangeQuery}(DB, dist, p, \varepsilon)$
    \If{$|N| > minPts$}
        \State Seed set $S$ = $\{p\}$
    \EndIf    
    \For{each $q$ in $S$}
        \If{$label(q) \neq undefined$}
        \EndIf
        \State $N \gets \textsc{RangeQuery}(DB, dist, q, \varepsilon)$
        \If{$|N| > minPts$}
            \State \;
            \State \;
            \State \;
            \State $label(q) \gets c$
            \State $S \gets S \cup N$
        \EndIf
    \EndFor
    \end{algorithmic}
\end{algorithm}
\end{minipage}
\begin{minipage}[t]{0.5\textwidth}
\begin{algorithm}[H]
    \centering
    \caption{Find a flat cluster by constructing MST on mutual reachability distance)}\label{algorithm2}
    \begin{algorithmic}[1]
        \State $N \gets \textsc{RangeQuery}(DB, dist_{\mathrm{mr}}, p, \varepsilon)$
        \State Seed set $S \gets \{p\}$
        \State Edge set $E \gets \emptyset$
        \While{1}
            \State $N \gets \emptyset$
            \For{each $s \in S$}
                \State $N \gets N \cup \textsc{RangeQuery}(DB, dist_{\mathrm{mr}}, s, \varepsilon)$
            \EndFor
            \If{$N =$ \textit{empty}}
                \State \textbf{break}
            \EndIf
            \State $(q, d) \gets \textsc{NearestLink}(S, N, dist_{\mathrm{mr}})$
            \State $label(q) \gets c$
            \State $S \gets S \cup \{q\}$
            \State $E \gets E \cup \{d\}$ 
        \EndWhile
    \end{algorithmic}
\end{algorithm}
\end{minipage}
\end{figure*}

\section{Methodology} \label{sec:method}
Section~\ref{sec:HCmethod} discusses the methodologies of two hierarchical clustering algorithms applied to analyse inner hierarchical clustering phenomena, SLINK~\cite{gower1969minimum, sibson1973slink} and HDBSCAN~\cite{campello2013density}, by examining their relationships. Section~\ref{sec:CCMfscore} presents the methodology of the proposed Hierarchical Cluster-Class Matching (HCCM) method used to interpret hierarchical representation clusters of each inner hierarchical phenomenon, with the F-score metric used to quantify the matching degree. Section~\ref{sec:CCMlscore} presents a new quantification metric, the Liebig score (L-score), for the HCCM method, which allows a more detailed diagnostic interpretation of the matching degree.


\subsection{Relationships of SLINK and HDBSCAN} \label{sec:HCmethod}


SLINK~\cite{gower1969minimum} is one of the most well-known hierarchical clustering algorithms used to analyse the hierarchical clustering phenomena present in the data, where each phenomenon represents a natural organisation of data as clusters with hierarchical relationships. In particular, this work uses SLINK to analyse representations from the speaker recognition network in order to uncover the hierarchical clustering phenomena within the representation space (i.e. inner hierarchical clustering phenomena). HDBSCAN is another well-known hierarchical clustering algorithm we used to analyse inner hierarchical clustering phenomena. It is the hierarchical extension of DBSCAN, in which flat data clusters produced by running DBSCAN at different search radii can be naturally nested to form hierarchical data clusters. This section aims to introduce these two algorithms by examining their relationship.

The relationship between HDBSCAN and SLINK lies in that their fast implementations both rely on the Minimum Spanning Tree (MST)~\cite{prim1957shortest, kruskal1956shortest}. Specifically, an MST is a tree structure connecting all data points with edges such that the total edge weight is minimised. In the fast implementation of SLINK, the MST is first constructed over the data points, and hierarchical data clusters can then be efficiently obtained by pruning the MST. In contrast, Campello et al.~\cite{campello2013density} proposed that the fast implementation of HDBSCAN can fully adopt the MST-based implementation of SLINK, replacing only the original distance metric with the mutual reachability distance metric they designed. To the best of our knowledge, the equivalence between HDBSCAN and SLINK has received little discussion in the literature. We now discuss this equivalence steps by steps.

Algorithm~\ref{algorithm1} and Algorithm~\ref{algorithm2} present side-by-side pseudocodes showing how a flat cluster of data points (i.e. network representations in our experiments) is analysed by DBSCAN and by constructing an MST using the mutual reachability distance metric, respectively. For Algorithm~\ref{algorithm1}, similar to the peoducode show in Fig.~\ref{fig:pesudo}, it first selects an start data point $p$ and determines whether it is a \textit{core point} that satisfies the density constraint, using the $\textsc{RangeQuery}$ function to find at least $minPts$ neighbours within the $\varepsilon$ search radius in the distance space defined by $dist$ distance metric. Algorithm~\ref{algorithm1} then iteratively selects new core points from the reachable neighbours of previously identified core points until no new core points can be found for a certain cluster. 



Furthermore, Algorithm~\ref{algorithm2} also begins from the same initial point $p$. Specifically, it invokes the $\textsc{RangeQuery}$ function to query the neighbours of point $p$ within the search radius $\varepsilon$, with the distance metric changed to the $dist_{\mathrm{mr}}$, which is the mutual reachability distance metric. In more detail, given two specific data points $x_1$ and $x_2$ (i.e. representations) in the dataset $DB$, their distance in the original distance space is $dist(x_1, x_2)$, while in the mutual reachability distance space it is defined as:

\begin{equation}
\begin{split}
dist_{\mathrm{mr}}(x_1, x_2) = \max \big( & dist_{\mathrm{core}}(x_1), \\
                                             & dist_{\mathrm{core}}(x_2), \\
                                             & dist(x_1, x_2) \big)
\end{split} \label{eq:mrd}
\end{equation}

\noindent where $dist_{\mathrm{core}}(x_i)$ is the minimal radius containing $minPts$ neighbours near $x_i$ in the original distance space defined by $dist$. Taking the maximum in Equation~\eqref{eq:mrd} ensures that if either $x_1$ or $x_2$ is not dense enough, their mutual reachability distance is inflated, so that only those points dense enough in the original distance space can be neighbours within the mutual reachability distance space. In the special case of $minPts = 0$, $dist_{\mathrm{core}}(x_i) = 0$ for all points, and the mutual reachability distance $dist_{\mathrm{mr}}(x_1, x_2)$ reduces to the original distance $dist(x_1, x_2)$, with no density constraint imposed.

Returning to line~1 of Algorithm~\ref{algorithm2}, all points returned by $\textsc{RangeQuery}(DB,\, {dist}_{\mathrm{mr}},\, p,\, \varepsilon)$ are guaranteed to be core points and stored in $S$, eliminating the explicit density check required in Algorithm~\ref{algorithm1}. Then, in the loop from line~4 to~16, $\textsc{RangeQuery}(DB,\, {dist}_{\mathrm{mr}},\, q,\, \varepsilon)$ is iteratively executed with each element $q \in S$ as the query point, and the newly found core points are collected in the set $N$. Next, the function $\textsc{NearestLink}(S,\, N,\, \mathrm{dist}_{\mathrm{mr}})$ identifies the point $q \in N$ closest to any point in $S$ under ${dist}_{\mathrm{mr}}$, returning $q$ along with its shortest distance $d$ to $S$. The core point $q$ is then labelled as a new cluster member and added to $S$, while distance $d$ is added to $E$. This process repeats until no new core points are added to $S$, at which point $S$ contains all members of a particular flat cluster and $E$ contains all smallest distances, both of which can be used to form the MST over cluster members.

Comparing Algorithm~\ref{algorithm2} and Algorithm~\ref{algorithm1}, Algorithm~\ref{algorithm2} expands cluster membership by selecting the nearest core point among all neighbours reachable from the current cluster members, borrowing the idea from Prim's algorithm~\cite{prim1957shortest}, whereas Algorithm~\ref{algorithm1} expands the cluster using any reachable core point among the neighbours of the current cluster members. This distinction only affects the order in which cluster members are added, but not the final set of cluster members~\cite{campello2013density}. Hence, running DBSCAN to find a flat cluster is equivalent to finding a flat cluster by constructing the MST in the mutual reachability distance space, when both are configured with the same setup (i.e. search radius $\varepsilon$, density parameter $minPts$, and starting point $p$).

The hierarchical data clusters of HDBSCAN naturally consist of flat clusters obtained by repeatedly running Algorithm~\ref{algorithm1} at different search radii $\varepsilon$ and different starting points $p$, the details of which are not further elaborated here. However, McInnes et al.~\cite{mcinnes2017hdbscan} proposed a more efficient HDBSCAN implementation by running Algorithm~\ref{algorithm2} with a sufficiently large search radius to construct an MST covering all data points. Then, flat data clusters at progressively lower radii can be obtained quickly by pruning this MST, where removing edges in decreasing order of distance corresponds to progressively reducing the search radius, efficiently yielding flat data clusters at each corresponding radius and forming the hierarchical data clusters. McInnes et al.'s MST-based HDBSCAN implementation is inspired by the MST-based SLINK implementation, which constructs and prunes the MST in the original distance space rather than the mutual reachability distance space. In particular, as discussed earlier, when $minPts = 0$, the mutual reachability distance space reduces to the original distance space, meaning that the MST-based HDBSCAN with $minPts = 0$ is equivalent to the MST-based SLINK.

\subsection{Hierarchical Cluster-Class Matching for interpretation}~\label{sec:CCMfscore}
In Section~\ref{sec:old_ccm_method}, CCM~\cite{rosenberg2007v} computes an overall matching score by finding the highest matching score between each predefined data division and all data clusters, implicitly identifying a best-matched cluster for each predefined division. We are therefore motivated to explicitly identify such best-matched cluster-class pairs in the context of interpreting hierarchical representation clusters produced by an algorithm (i.e. SLINK or HDBSCAN). The more predefined semantic classes can divide representations, the more hierarchical representation clusters can be correspondingly best matched to these predefined representation divisions. Such best-matched clusters can then be interpreted as the semantic classes corresponding to those predefined representation divisions. To this end, we develop the Hierarchical Cluster-Class Matching (HCCM) method.

\begin{figure}
    \centering
    \includegraphics[width=0.97\linewidth]{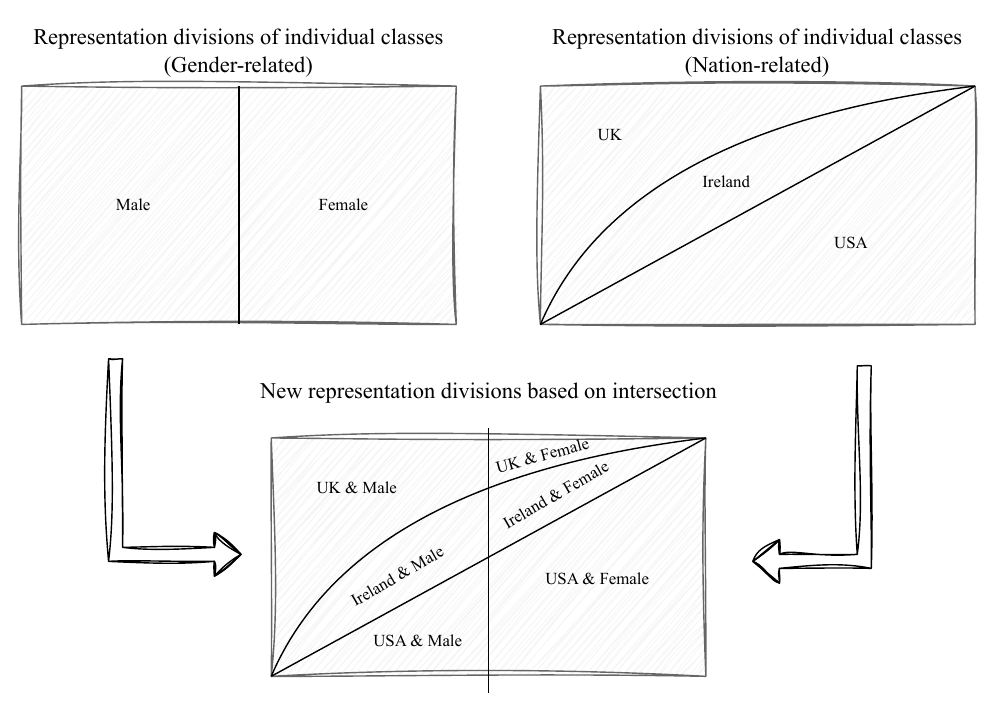}
    \caption{An illustration of how the predefined representation divisions of two gender-related individual semantic classes and three nationality-related individual semantic classes are intersected to form the predefined representation divisions of conjunctive semantic classes.}
    \label{fig:conjunctive_semantic_class}
\end{figure}

We now discuss how representations can be divided based on diverse semantic classes in HCCM. Specifically, given a set of model inputs (e.g.\ spectrograms of utterances) and their representations extracted from a speaker recognition neural network (typically from the penultimate layer), the speaker identity of each model input is known since the task is speaker recognition. Suppose the entire dataset involves $n_\mathrm{spk}$ distinct speaker classes; all model inputs can be divided into $n_\mathrm{spk}$ classes, and their corresponding representations are divided in the same manner, forming $n_\mathrm{spk}$ predefined representation divisions. In HCCM, users can additionally annotate each model input with attribute classes beyond speaker identity, such as speaker nationality and gender, thereby introducing more ways to divide representations. As illustrated in Fig.~\ref{fig:conjunctive_semantic_class} (for example purposes only), the top-left block shows that all representations can be divided by two gender-related classes, and the top-right block shows that all representations can be divided by three nationality-related classes.

Building on this, to create more predefined representation divisions, HCCM additionally applies intersection operations to the known predefined representation divisions, thereby generating new predefined representation divisions. As shown in the lower blocks of Fig.~\ref{fig:conjunctive_semantic_class}, taking the pairwise intersections of 2 gender-related representation divisions and 3 nationality-related representation divisions yields $2 \times 3 = 6$ new representation divisions, where each division contains representations belonging simultaneously to a specific nationality class and a specific gender class. HCCM defines an atomic semantic class that cannot be further decomposed as an \textit{individual semantic class}; for example, `UK' is an individual semantic class that determines a non-intersected representation division. HCCM defines a semantic class formed by combining two or more individual semantic classes through a conjunctive operation as a \textit{conjunctive semantic class}; for example, `UK $\wedge$ male' or `UK \& male' is a conjunctive semantic class, where `$\wedge$' and `\&' denote the Boolean conjunction operator, which can also be expressed in natural language as `UK and male', that determines an intersected representation division.

Mathematically, the predefined representation divisions of individual semantic classes can be denoted as $C = \{ c_1, \dots, c_{n_{\mathrm{class}}} \}$, where $n_{\mathrm{class}}$ is the total number of distinct individual semantic classes, spanning different attributes such as speaker nationality, gender, and identity. $c_i$ denotes the predefined representation division of the $i$-th individual semantic class, containing the indices of those representations pre-labelled with the $i$-th individual semantic class. In addition, predefined representation divisions of conjunctive semantic classes can be denoted as follows:

\begin{equation}
\begin{aligned}
\mathcal{C}
  = \bigl\{\, c_T \mid T \subseteq I,\; |T| = 2,\; c_T \neq \varnothing \bigr\}, \quad
c_T = \bigcap_{t \in T} c_t.
\end{aligned}
\label{eq:conj_class}
\end{equation}

\noindent Here, $I = \{1, \dots, n_{\mathrm{class}}\}$ denotes the index set of all individual semantic classes. Equation~\eqref{eq:conj_class} enumerates all possible pairs of individual semantic classes from $I$ as $T$. $c_T$ is the predefined representation division of the conjunctive semantic class corresponding to $T$, obtained by taking the intersection $\bigcap_{t \in T} c_t$, which contains the indices of the representations simultaneously both of each individual semantic class $t$ in $T$. If the two selected classes in $T$ are disjoint, $c_T$ is empty and no conjunctive class is formed. In practice, not all pairwise combinations yield non-empty intersections, so the actual number of valid conjunctive classes is fewer than the theoretical maximum of $\binom{n_{\mathrm{class}}}{2}$. Regardless, $\mathcal{C}$ collects the predefined representation divisions of all valid conjunctive semantic classes.

HCCM then uses all predefined representation divisions (i.e. both $C$ and $\mathcal{C}$) to perform pairwise matching with hierarchical representation clusters produced by a certain algorithm. In particular, given hierarchical clusters produced by applying a hierarchical clustering algorithm (e.g.\ SLINK or HDBSCAN) to the representations, these hierarchical clusters are denoted as $K = \{ k_1, \dots, k_{n_{\mathrm{cluster}}} \}$. Here, $n_{\mathrm{cluster}}$ represents the total number of hierarchical representation clusters identified by the algorithm, which is typically far smaller than $n_{\mathrm{class}}$. $k_j$ denotes the $j$-th hierarchical representation cluster, containing the union of indices of all flat representation clusters that compose it, with no duplicate indices. Lastly, the pairwise matching process is formulated as follows:

\begin{equation} \label{eq:HCCM_f}
\begin{aligned}
& \mathbb{C}^{(0)} &&= C \cup \mathcal{C},  \\
& (c^*_l, k^*_l) &&= \arg\max_{c \in \mathbb{C}^{(l-1)},\, k \in K} f_{\mathrm{score}}(c, k), \\
& \mathbb{C}^{(l)} &&= \mathbb{C}^{(l-1)} \setminus \{ c^*_l \},  l = 1, \dots, |\mathbb{C}^{(0)}|
\end{aligned}
\end{equation}

\noindent Here, Equation~\eqref{eq:HCCM_f} describes an iterative matching process that finds a best-matched hierarchical representation cluster for each predefined representation division. Initially, $\mathbb{C}^{(0)}$ includes all predefined representation divisions of both individual and conjunctive semantic classes. At each iteration $l$, a certain best-matched cluster-class pair $(c^*_l, k^*_l)$ is identified by selecting the pair that achieves the highest F-score among all remaining candidate predefined representation divisions in $\mathbb{C}^{(l-1)}$ and all hierarchical representation clusters in $K$. The predefined representation division $c^*_l$ is then removed from $\mathbb{C}^{(l-1)}$, ensuring that each predefined representation division is matched to at most one hierarchical representation cluster. This process repeats until all predefined representation divisions have been matched, yielding a total of $|\mathbb{C}^{(0)}|$ best-matched cluster-class pairs. Each pair interprets a hierarchical representation cluster using the semantic class corresponding to its best-matched predefined representation division. 

Eventually, HCCM can provide a richer semantic interpretation of each inner hierarchical clustering phenomenon when more individual semantic classes are available and more conjunctive semantic classes can be constructed from them to interpret its hierarchical representation clusters.

\subsection{Liebig's score} \label{sec:CCMlscore}


Once HCCM identifies the best-matched cluster-class pairs, we believe that further understanding how well each match is achieved is important, including whether the match is perfect or the reason for any mismatch. Section~\ref{sec:CCMfscore} quantifies the matching degree of each best-matched cluster-class pair using the F-score metric; however, as argued in Section~\ref{sec:old_ccm_method}, the F-score lacks a clear and meaningful value interpretation, unlike precision and recall~\cite{christen2023review}, let alone diagnosing the mismatch for each cluster-class pair according to an F-score-based matching degree. To address this, we propose a new metric, Liebig's score (L-score), as a replacement for the F-score to quantify the matching degree for each cluster-class pair, which can offer a diagnostic interpretation of each match.

The L-score is inspired by Liebig's Law of the Minimum~\cite{de1994liebig}, a principle stating that the overall performance of a system is determined by its weakest component. Analogously, we assume that the matching performance between a representation cluster and a predefined representation division is primarily limited by the weakest matching factor. In the context of a cluster-class pair, precision and recall are treated as the two matching factors that characterise the matching performance. Based on Liebig's Law, the L-score determines the matching degree of a cluster-class pair as the smaller of its precision and recall values, i.e.\ the most limiting matching factor. Formally, the L-score is calculated as follows:


\begin{equation}
\begin{split}
l_{\mathrm{score}}(c, k) = \min\left(\frac{|c \cap k|}{|c|}, \frac{|c \cap k|}{|k|}\right)  \\
                = \frac{|c \cap k|}{\max(|c|,\: |k|)} \label{eq: c-score1}
\end{split}
\end{equation}


\begin{figure*}[t]
    \centering
    \includegraphics[width=0.95\linewidth]{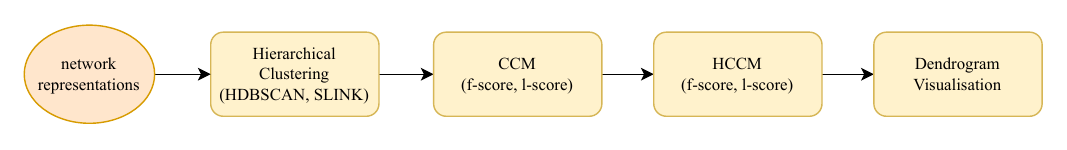}
    \caption{An overview of experimental procedures}
    \label{fig:EXP_procedure}
\end{figure*}

\noindent Here, $\frac{|c \cap k|}{|k|}$ is the precision for matching representation cluster $k$ with a predefined representation division $c$. We define precision in this context as the proportion of representations in cluster $k$ that truly belong to its matched semantic class (i.e.\ are pre-labelled with the corresponding semantic class). We define recall (i.e.\ $\frac{|c \cap k|}{|c|}$) as the proportion of representations within the predefined representation division $c$ that are correctly assigned to its matched cluster $k$ (i.e.\ retrieved by the clustering algorithm). Equation~\eqref{eq: c-score1} selects the minimum of the precision and recall values as the L-score, representing the matching degree of the cluster-class pair $(k, c)$.


We now demonstrate how to diagnostically interpret the match of each cluster-class pair based on the L-score-based matching degree, as it offers direct access to the most limiting matching factor. Suppose the L-score-based matching degree for a given cluster-class pair $(k, c)$ is $0.1$. Given our assumption drawn from Liebig's Law of the Minimum, its matching performance is limited by either a precision of $0.1$ or a recall of $0.1$. If precision is the smaller value, the $0.1$ L-score-based matching degree is interpreted as: the matching performance is limited by the fact that $90\%$ of the representations within the cluster do not belong to the matched semantic class, or alternatively, only $10\%$ of them truly belong to the matched semantic class. If recall is the smaller value, the matching degree is interpreted as: the mismatch is due to the fact that $90\%$ of the representations in the predefined representation division are not retrieved by the cluster, or alternatively, only $10\%$ of them are successfully retrieved.

\section{Experimental Procedures and Setups} 


Fig.~\ref{fig:EXP_procedure} provides a visual overview of our experimental procedures, while this section details the setup for each procedure.

\textbf{Preparing network representations}: Our experiments begin with obtaining network representations for subsequent procedures. Specifically, the speaker recognition network trained by Chung et al.~\cite{chung2018voxceleb2} is downloaded from \href{https://github.com/clovaai/voxceleb_trainer}{https://github.com/clovaai/voxceleb\_trainer}. This network is based on the ResNet34 architecture~\cite{he2016deep}, trained with the prototypical contrastive loss~\cite{snell2017prototypical, chung2020defence} on mel spectrograms of 2-second audio clips from the VoxCeleb2 dataset~\cite{chung2018voxceleb2}, with an Equal Error Rate (EER) of $2.18\%$ evaluated on the speaker verification task using the VoxCeleb1 test set~\cite{nagrani2017voxceleb}. Speaker embeddings (i.e.\ representations) are then extracted from the penultimate layer of this trained network while classifying mel spectrograms of audio clips from the VoxCeleb1 test set. Specifically, mel spectrograms of 20, 100, 200, and 400 frames are used as inputs, corresponding to audio clips of 0.2, 1, 2, and 4 seconds in duration, respectively.


\textbf{Hierarchical clustering}: In Fig.~\ref{fig:EXP_procedure}, SLINK and HDBSCAN are applied to the representations. Specifically, the representations are categorised into four groups based on the audio duration (i.e.\ 0.2-, 1-, 2-, and 4-second audio clips), to each of which SLINK and HDBSCAN are applied separately. HDBSCAN is implemented by McInnes et al.~\cite{mcinnes2017hdbscan}, available at \href{https://github.com/scikit-learn-contrib/hdbscan}{https://github.com/scikit-learn-contrib/hdbscan}. McInnes et al.'s implementation is based on the MST-based SLINK algorithm operating on the mutual reachability distance space, which is derived from the Euclidean distance space. We therefore do not implement SLINK separately, as setting $minPts = 0$ reduces the mutual reachability distance space to the original Euclidean distance space, making McInnes et al.'s implementation equivalent to running MST-based SLINK directly in the original Euclidean distance space. More detailed setups of the hierarchical clustering algorithms are discussed in Section~\ref{sec:result}.


\textbf{CCM and HCCM}: As illustrated in Fig.~\ref{fig:EXP_procedure}, given the hierarchical clustering results (i.e.\ hierarchical representation clusters) obtained from applying SLINK and HDBSCAN to representations of different audio durations, CCM~\cite{rosenberg2007v} is first performed, followed by HCCM. In particular, each hierarchical clustering result corresponds to a certain inner hierarchical clustering phenomenon of our examined speaker recognition network, and CCM evaluates how well each phenomenon is analysed by the algorithm, reflecting the quality of its hierarchical representation clusters. The phenomenon achieving the optimal CCM evaluation score is then semantically interpreted by the proposed HCCM method.

The predefined representation divisions used in both CCM and HCCM are derived from the data annotations of the VoxCeleb1 test set~\cite{nagrani2017voxceleb}, in which all utterances and their corresponding representations can be divided into 40 identity-related, 2 gender-related, and 12 nationality-related individual semantic classes. Additionally, 24 conjunctive semantic classes are constructed by intersecting the predefined representation divisions of the 2 gender-related individual classes with those of the 12 nationality-related individual classes, yielding a total of $40 + 2 + 12 + 24 = 78$ predefined representation divisions. For HCCM, all 78 predefined representation divisions are directly used jointly for pairwise matching. For CCM, predefined representation divisions of each attribute type are separately used to evaluate the outcomes of SLINK and HDBSCAN, with no consideration of whether these representation divisions are individual or conjunctive as categorised in HCCM; for example, predefined representation divisions related to gender, including both individual and conjunctive gender-related classes, are used together to evaluate gender-related alignment. Lastly, the matching degree of each cluster-class pair in HCCM is quantified using the L-score metric, whereas the overall matching degree in CCM is quantified using both the F-score and L-score metrics.


\textbf{Dendrogram visualisation}: The dendrogram visualisation is performed as the last procedure, as shown in Fig.~\ref{fig:EXP_procedure}. In particular, we visualise the inner hierarchical clustering phenomenon achieving the highest CCM evaluation score as a dendrogram using the code provided by McInnes et al.~\cite{mcinnes2017hdbscan}, and the best-matched cluster-class pairs yielded by HCCM with L-score-based matching degrees of at least $0.25$ are empirically selected and manually annotated on this dendrogram.

\section{Result analysis} \label{sec:result}

\begin{figure*}[ht]
    \centering
    \begin{subfigure}[b]{\textwidth}
        \centering
        \includegraphics[width=\linewidth]{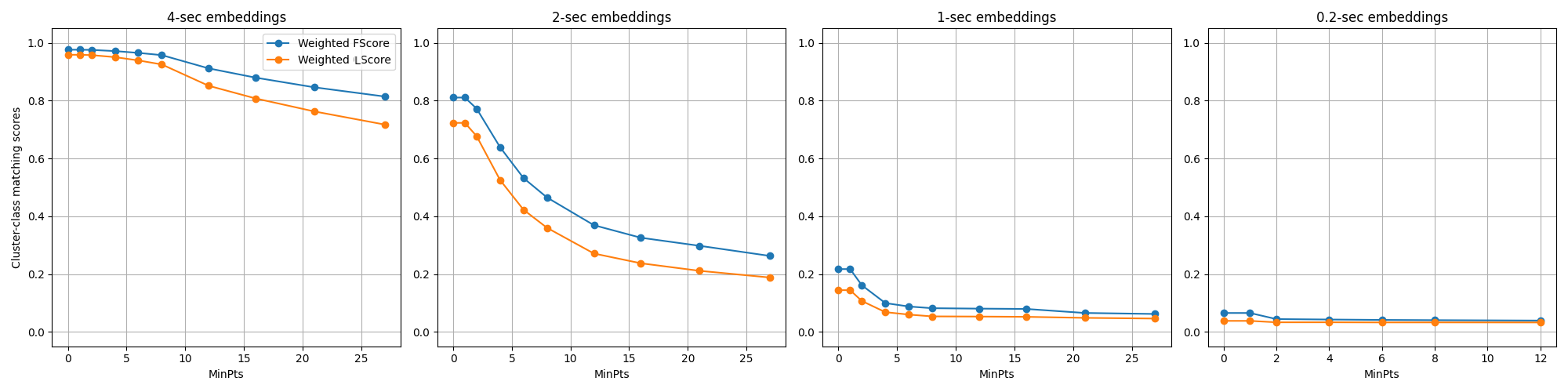}
        \caption{Matching identity-related classes}
        \label{fig:identity}
    \end{subfigure}
    \begin{subfigure}[b]{\textwidth}
        \centering
        \includegraphics[width=\linewidth]{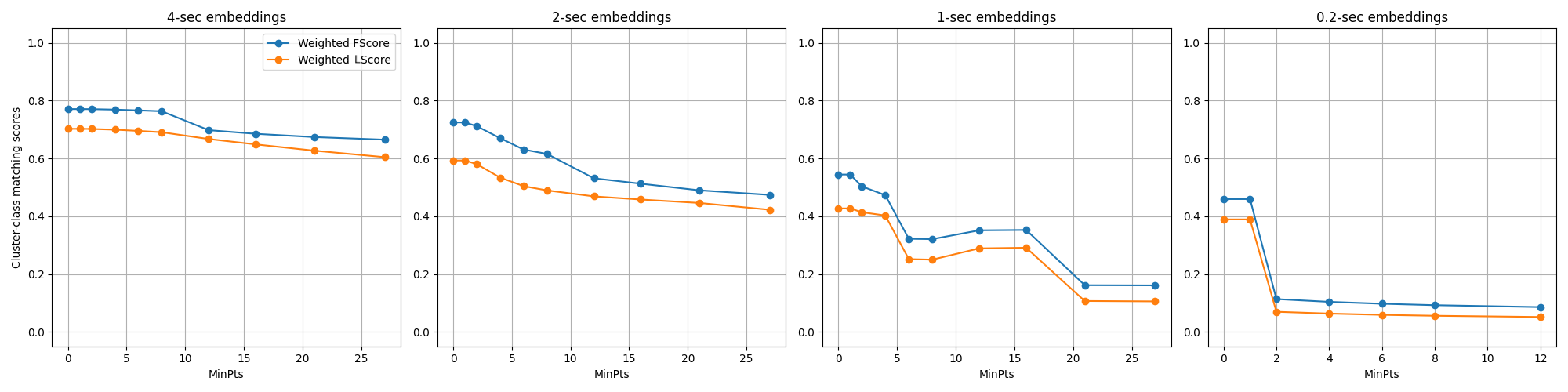}
        \caption{Matching nationality-related classes}
        \label{fig:nation}
    \end{subfigure}
    \begin{subfigure}[b]{\textwidth}
        \centering
        \includegraphics[width=\linewidth]{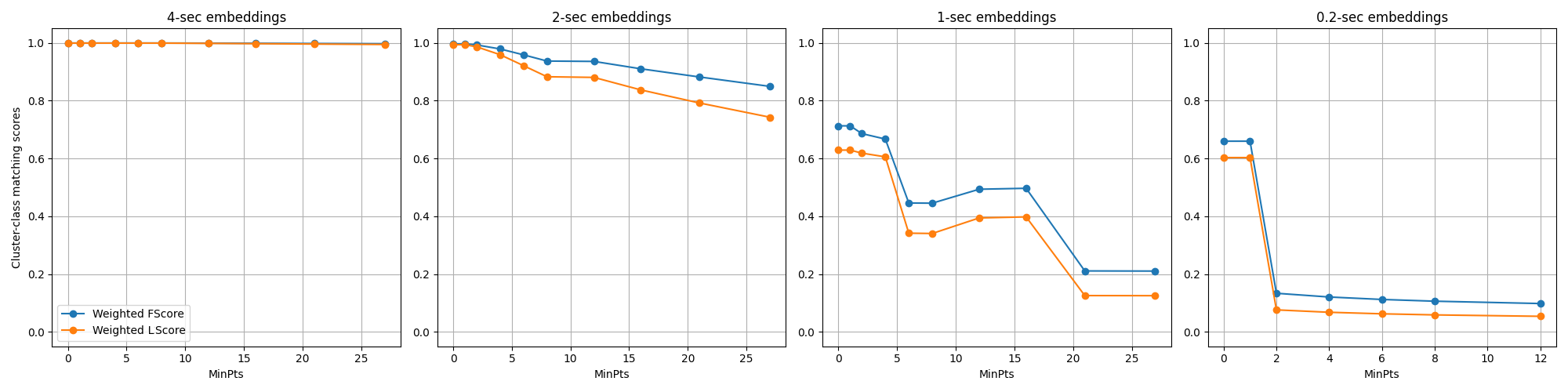}
        \caption{Matching gender-related classes}
        \label{fig:gender}
    \end{subfigure}
    \caption{Cluster-class matching degrees~\cite{rosenberg2007v} for evaluating the hierarchical representation clusters obtained by applying SLINK (i.e. \(minPts = 0\)) and HDBSCAN (i.e. \(minPts = 2, 4, 6, 8, 12, 16, 21, 27)\)) to representations (i.e. embeddings) of \(0.2\), \(1\), \(2\), and \(4\)-second audios.}
    \label{fig:clustering_result}
\end{figure*}

This section presents results for analysing, visualising, and interpreting hierarchical clustering phenomena within the representation space (i.e.\ inner hierarchical clustering phenomena) of the speaker recognition network, with the initial question of what unknown organisations exist in the network's representation space. In particular, Section~\ref{sec:CCMresult} presents how well HDBSCAN and SLINK analyse different inner hierarchical clustering phenomena, as evaluated using the CCM method~\cite{rosenberg2007v}. Section~\ref{sec:xdendrogram} presents the dendrogram visualisation and HCCM's semantic interpretation of the best-analysed inner hierarchical clustering phenomenon revealed by either SLINK or HDBSCAN.

\subsection{Analysing inner hierarchical clustering: CCM's evaluation} \label{sec:CCMresult}

CCM~\cite{rosenberg2007v} is used to evaluate how well SLINK~\cite{gower1969minimum} and HDBSCAN~\cite{campello2013density} analyse inner hierarchical clustering phenomena of the speaker recognition network, by scoring the quality of the hierarchical representation clusters corresponding to each phenomenon. Specifically, higher-quality hierarchical representation clusters are expected to overall align more closely with the predefined grouping information of the representations.


Fig.~\ref{fig:identity} presents the CCM overall matching degrees between hierarchical representation clusters produced by SLINK and HDBSCAN under different setups and the predefined representation divisions of identity-related semantic classes. The four subplots correspond to representations extracted from 4-, 2-, 1-, and 0.2-second audio clips, respectively. Within each subplot, the $x$-axis represents the $minPts$ parameter, where $minPts = 0$ corresponds to SLINK and $minPts > 0$ corresponds to HDBSCAN. The $y$-axis represents the overall matching degree, with blue and yellow points denoting the F-score and L-score metrics, respectively.

As shown in Fig.~\ref{fig:identity}, the CCM matching degrees quantified by both the L-score and F-score metrics gradually decline as $minPts$ increases, indicating that HDBSCAN imposing a stronger density requirement to produce hierarchical representation clusters cannot improve their alignment with the predefined grouping information of speaker identity. The highest overall matching degree is obtained when $minPts = 0$ on 4-second audio representations, achieving a value of nearly $1.0$. This suggests that applying SLINK on the Euclidean space to the longest audio representations yields hierarchical representation clusters that best align with the predefined representation divisions of identity-related semantic classes, meaning that the corresponding inner hierarchical clustering phenomenon best preserves the grouping information related to speaker identities.


Furthermore, Fig.~\ref{fig:nation} and Fig.~\ref{fig:gender} present the CCM matching degrees for nationality-related and gender-related semantic classes, respectively. The optimal hierarchical representation clusters in both figures are consistently produced by applying SLINK to 4-second audio representations, achieving an overall matching degree exceeding $0.6$ for nationality-related representation divisions and nearly $1.0$ for gender-related representation divisions. Taken together, the inner hierarchical clustering phenomenon analysed by SLINK on 4-second representations is semantically meaningful, as the corresponding hierarchical representation clusters preserve the strongest alignment with the predefined grouping information related to gender, nationality, and identity. It is worth noting that the inner hierarchical clustering phenomena analysed under other configurations are not entirely arbitrary either, as their hierarchical representation clusters still retain a non-negligible degree of alignment with the predefined grouping information.

\subsection{Visualising and interpreting inner hierarchical clustering: An annotated dendrogram} \label{sec:xdendrogram}


This section visualises the best-analysed inner hierarchical clustering phenomenon (i.e.\ the hierarchical representation clusters obtained by applying SLINK to 4-second representations) as a dendrogram. To further deepen our understanding of this phenomenon, the proposed HCCM method is applied to semantically interpret the clusters shown in the dendrogram, given that Section~\ref{sec:CCMresult} has demonstrated that the predefined grouping information of semantic classes is preserved within this phenomenon, yet the specific hierarchical position of each semantic class remains unknown. HCCM achieves this by performing pairwise matching between these clusters and the predefined representation divisions of both individual and conjunctive semantic classes.

Fig.~\ref{fig:dendrogram_800} depicts the dendrogram visualised using the optimal hierarchical representation clusters produced by applying MST-based SLINK to 4-second audio representations. Since the number of these clusters reaches the order of thousands, only clusters containing more than 800 representations are retained for visualisation, to keep the dendrogram concise and clear. The $y$-axis represents $\lambda = \tfrac{1}{\varepsilon}$, where $\varepsilon$ is the search radius used both to construct the MST in SLINK and to prune it for obtaining hierarchical representation clusters.

Each icicle-shaped object in Fig.~\ref{fig:dendrogram_800} represents a hierarchical representation cluster, which may be understood as the aggregation of a series of flat representation clusters across a continuous range of $\lambda$ values. At any fixed $\lambda$, pruning the MST at distance threshold $\varepsilon$ can be seen as drawing a horizontal cross-section at that $\lambda$ in the dendrogram, where the overlapping segments between the cross-section and the icicles correspond to the flat representation clusters obtained at that $\lambda$. The length of each overlapping segment is determined by the number of representations in the corresponding flat representation cluster. As $\lambda$ increases (i.e.\ $\varepsilon$ decreases), the edges in the MST linking representations of different hierarchical clusters are progressively removed. This either reduces the number of representations in a hierarchical cluster, manifested as the shrinking or vanishing of an icicle in the dendrogram, or causes a larger hierarchical cluster to break into smaller ones, manifested as a larger icicle splitting into smaller ones.

Furthermore, the HCCM's pairwise matching results in Fig.~\ref{fig:dendrogram_800} are annotated using icons and text labels. Specifically, each icon is placed adjacent to a hierarchical representation cluster to indicate the semantic class whose predefined representation division HCCM identifies as the best match for that cluster. Two types of icons are used: icons without the \& symbol denote individual semantic classes, while icons with the \& symbol denote conjunctive semantic classes. The meaning of each icon can be found in the three legends provided alongside the dendrogram. Additionally, a text label near each icon displays the L-score-based matching degree of the corresponding best-matched cluster-class pair. When the L-score selects recall as the limiting factor, the label is displayed as \texttt{rec:}\ followed by the recall value; when precision is selected, it is displayed as \texttt{pre:}\ followed by the precision value. To keep the dendrogram concise, only icons and text labels corresponding to best-matched cluster-class pairs achieving an L-score-based matching degree of at least $0.25$ are displayed in Fig.~\ref{fig:dendrogram_800}.

\afterpage{
\clearpage
\begin{landscape}
\begin{figure}
   \centering
   \includegraphics[width=650pt]{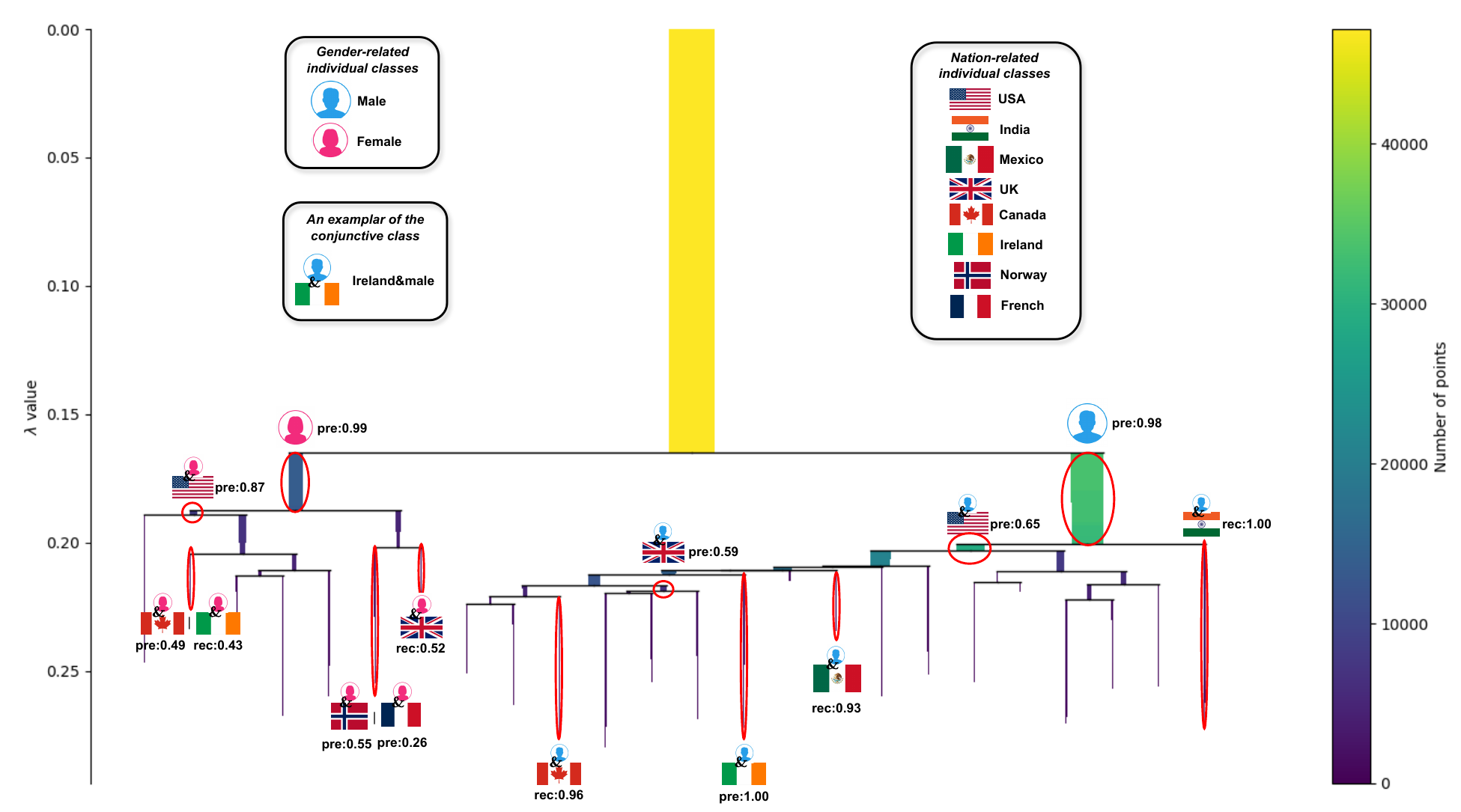}
  \caption{Visualising hierarchical representation clusters produced by applying SLINK to 4-second speaker embeddings as a dendrogram, with icon annotations showing the semantic interpretations that HCCM offered for these representation clusters, and text labels showing the L-score-based matching degree that HCCM measured for each best-matched cluster-class pair.~\label{fig:dendrogram_800}}
\end{figure}
\end{landscape}
\clearpage
}

Observing Fig.~\ref{fig:dendrogram_800}, there is a large icicle at the top of the dendrogram, representing the root hierarchical representation cluster. As $\lambda$ increases, the width of this icicle diminishes, indicating that fewer representations remain in this root cluster. At a certain point, the root hierarchical cluster splits into two sibling hierarchical clusters, interpreted by HCCM as the male and female individual classes, respectively. Based on the text annotations, the L-score identifies precision as the limiting factor for both matching cases: the matching performance of the male cluster is limited by the fact that $2\%$ of representations within it do not belong to the male class, while that of the female cluster is limited by $1\%$ not belonging to the female class. In the following paragraphs, we observe the interpretations that HCCM offers for the subtrees derived from these two sibling clusters.


\textbf{Underneath the male representation cluster}, the male representation cluster further splits into two sibling hierarchical clusters as $\lambda$ increases, interpreted by HCCM as the conjunctive semantic classes of India\&male and of USA\&male, respectively. For the USA\&male representation cluster, the L-score selects precision as the limiting factor, and a diagnostic interpretation can be given as follows: the matching performance is limited by the fact that only $65\%$ of the representations within the cluster belong to both the male and USA classes simultaneously (i.e.\ the conjunctive class of USA\&male), meaning that the remaining $35\%$ imprecisely belong to other semantic classes (e.g.\ males from other countries).

\textbf{Underneath the USA\&male representation cluster.} The USA\&male representation cluster successively splits into sub-clusters interpreted as the Mexico\&male, Ireland\&male, UK\&male, and Canada\&male conjunctive semantic classes. Interestingly, the Ireland\&male, UK\&male, and Canada\&male representation clusters can be merged into a single higher-level representation cluster, suggesting that the underlying patterns inherent in male voices of these three countries share some similarity captured by the speaker recognition network. This similarity may be attributed to the historical and linguistic ties among Ireland, the UK, and Canada.

For the UK\&male representation cluster, the L-score selects precision as the limiting factor with a value of $0.59$, meaning that the matching performance is primarily constrained by the fact that only $59\%$ of representations within the cluster belong to both the UK and male individual classes, with the remaining $41\%$ belonging to other semantic classes. For the Canada\&male representation cluster, the L-score selects recall as the limiting factor with a value of $0.96$, meaning that only $96\%$ of all representations in the predefined representation division of the Canada\&male class are retrieved by the matched cluster. The Ireland\&male representation cluster achieves a perfect L-score of $1.00$, with both precision and recall at $1.00$, indicating that the cluster precisely and completely captures all representations belonging to the Ireland\&male conjunctive class.

\textbf{Underneath the female representation cluster.} Underneath the female representation cluster, one hierarchical representation cluster at a high-level position is best matched by the predefined representation division of the USA\&female conjunctive class. Interestingly, unlike the UK\&male representation cluster, which resides underneath the USA\&male representation cluster, the UK\&female representation cluster does not reside underneath the USA\&female representation cluster. This implies that the underlying patterns of USA\&male voices subsume those of UK\&male voices as captured by the speaker recognition network, whereas the underlying patterns of USA\&female and UK\&female voices are sufficiently distinct to remain separate clusters. For the UK\&female representation cluster, the L-score selects recall as the limiting factor with a value of $0.52$, meaning that $48\%$ of all representations pre-labelled as the UK\&female conjunctive class are not captured by the matched cluster.

The sibling hierarchical cluster of the UK\&female representation cluster serves as the best match for both the France\&female and Norway\&female conjunctive classes. If interpreted as the France\&female cluster, a considerable number of representations belonging to the Norway\&female class are imprecisely included; conversely, if interpreted as the Norway\&female cluster, a considerable number of representations belonging to the France\&female class are imprecisely included. It is therefore reasonable to interpret this cluster as a union of the France\&female conjunctive class and the Norway\&female conjunctive class. Furthermore, the Canada\&female and Ireland\&female conjunctive classes also share the same best-matched representation cluster, suggesting that this cluster could be interpreted as a union of the Ireland\&female and Canada\&female conjunctive classes.

In summary, the HCCM's semantic interpretation reveals the hierarchical positions of semantic classes within the SLINK-analysed inner hierarchical clustering phenomenon: gender-related individual classes (i.e.\ male and female) occupy the highest hierarchical position, while conjunctive classes combining nationality and gender information predominantly emerge as sub-clusters underneath. The matching degree of each best-matched cluster-class pair is diagnosed, identifying whether the mismatch is primarily due to representations not being retrieved or not precisely belonging to the matched class. Moreover, some representation clusters interpreted as conjunctive semantic classes can be merged or split, reflecting not only similarities or distinctions in how the network organises these representation clusters, but more meaningfully, similarities or distinctions among the semantic classes that interpret these representation clusters. These findings suggest that the significance of analysing, visualising, and interpreting inner hierarchical clustering phenomena extends beyond understanding the network's representation organisation itself, while also offering a lens to understand the data semantics with reference to such organisation.



\section{Discussion}

As discussed in Section~\ref{sec:CCMresult}, hierarchical representation clusters produced by applying SLINK or HDBSCAN to 4-second audio representations yield better CCM scores than those produced from shorter audio representations. This may be attributed to the model being better at processing longer audio inputs; when evaluating audio clips from the VoxCeleb1 test set, the average EERs for 4-, 2-, 1-, and 0.2-second clips are $2.17\%$, $2.35\%$, $2.98\%$, and $22.22\%$, respectively, indicating that the model generalises more reliably on longer audio. The consistency between this length-dependent generalisation ability and the corresponding CCM scores suggests that a better inner hierarchical clustering phenomenon can be analysed when the model exhibits greater generalisation capability.

\section{Conclusion}

Motivated by the question of how neural networks organise representations, this work proposes to analyse, visualise, and interpret inner hierarchical clustering phenomena, i.e.\ the different ways in which network representations naturally form clusters with hierarchical relationships. SLINK and HDBSCAN are separately applied to analyse how representations from the speaker recognition network form hierarchical representation clusters. The CCM method identifies the optimal inner hierarchical clustering phenomenon as that analysed by SLINK on the longest audio representations, as its hierarchical clusters overall align most closely with predefined grouping information related to speaker gender, nationality, and identity.

Furthermore, we visualise the optimal inner hierarchical clustering phenomenon as a dendrogram, with its hierarchical clusters semantically interpreted by the proposed HCCM method. HCCM reveals that representation clusters at upper hierarchical positions of the dendrogram best match predefined representation divisions of two gender-related individual semantic classes, while those at lower levels predominantly best match predefined representation divisions of conjunctive semantic classes combining gender and nationality information. Each best-matched cluster-class pair interprets a representation cluster as a semantic class, and the matching performance of each pair is diagnosed under the Liebig score metric. Importantly, the merging and splitting of clusters in the dendrogram also demonstrates the merging and splitting of the semantic classes that interpret them, thereby offering a pathway to understand data semantics through the network's inner hierarchical phenomena.

In future work, we will invite human experts in fields such as psychology~\cite{solso2005cognitive, gergen1973social}, linguistics~\cite{trager1958paralanguage}, speech pathology~\cite{travis1931speech}, and vocal pedagogy~\cite{heidemann2016system, xu2022paralinguistic} to interpret inner hierarchical clustering phenomena of different speaker recognition networks using more semantic classes, providing deeper insights into how more audio semantics can be understood through the lens of speaker recognition networks.

\section*{Acknowledgment}
Thanks to J.S.\ Chung and Arsha Nagrani for releasing the VoxCeleb dataset and for publishing the well-trained speaker recognition neural network. Thanks to researchers who have contributed to the deep learning and knowledge discovery methods used in this work. Thanks to everyone who remains sceptical of AI systems, and to those who stay curious about the unknown of this world.
\ifCLASSOPTIONcaptionsoff
  \newpage
\fi



\bibliographystyle{IEEEtran}
\bibliography{bibtex/bib/IEEEexample}
\end{document}